\documentclass[aps,prd,twocolumn,superscriptaddress]{revtex4-2}

\usepackage{graphicx}
\usepackage{dcolumn}
\usepackage{bm}
\usepackage{hyperref}
\usepackage{amsmath}
\usepackage{amssymb}
\usepackage{color}
\usepackage{makecell}

\begin{document}

\title{Scale-invariant Schr\"{o}dinger geometry in symmetric teleparallel gravity}

\author{Wei-Zhi Liu}
\email{WeizhiLiu@m.scnu.edu.cn}
\affiliation {Key Laboratory of Atomic and Subatomic Structure and Quantum Control (Ministry of Education), Guangdong Basic Research Center of Excellence for Structure and Fundamental Interactions of Matter, School of Physics, South China Normal University, Guangzhou 510006, China} 
\affiliation {Guangdong Provincial Key Laboratory of Quantum Engineering and Quantum Materials, Guangdong-Hong Kong Joint Laboratory of Quantum Matter, South China Normal University, Guangzhou 510006, China}

\author{Jing-Ying Wang}
\affiliation {Key Laboratory of Atomic and Subatomic Structure and Quantum Control (Ministry of Education), Guangdong Basic Research Center of Excellence for Structure and Fundamental Interactions of Matter, School of Physics, South China Normal University, Guangzhou 510006, China} 
\affiliation {Guangdong Provincial Key Laboratory of Quantum Engineering and Quantum Materials, Guangdong-Hong Kong Joint Laboratory of Quantum Matter, South China Normal University, Guangzhou 510006, China}

\author{Shi-Dong Liang}
\email{stslsd@mail.sysu.edu.cn}
\affiliation{School of Physics,
Sun Yat-Sen University, Guangzhou 510275, People’s Republic of China}

\author{Hong-Hao Zhang}
\email{zhh98@mail.sysu.edu.cn}
\affiliation{School of Physics,
Sun Yat-Sen University, Guangzhou 510275, People’s Republic of China}

\author{Tiberiu Harko}
\email{tiberiu.harko@aira.astro.ro}
\affiliation{Department of Physics, Babe\c s-Bolyai University, 1 Kog\u alniceanu Street, Cluj Napoca 400084, Romania,} 
\affiliation{Astronomical Observatory, 19 Cire\c silor Street, Cluj-Napoca 400487, Romania}

\author{Lei Ming}
\email{minglei@scnu.edu.cn}
\affiliation {Key Laboratory of Atomic and Subatomic Structure and Quantum Control (Ministry of Education), Guangdong Basic Research Center of Excellence for Structure and Fundamental Interactions of Matter, School of Physics, South China Normal University, Guangzhou 510006, China} 
\affiliation {Guangdong Provincial Key Laboratory of Quantum Engineering and Quantum Materials, Guangdong-Hong Kong Joint Laboratory of Quantum Matter, South China Normal University, Guangzhou 510006, China}

\date{\today}

\begin{abstract}
We construct a locally scale-invariant formulation of Schr\"{o}dinger geometry in symmetric teleparallel gravity. The unique scale transformation preserving autoparallelism, torsionlessness and the Schr\"{o}dinger affine structure is first identified, from which a quadratic scale-invariant action is obtained. Using the Palatini formalism, we show that the conditions required for scale invariance are exactly those ensuring that the affine connection is dynamically reduced to the Schr\"{o}dinger form. Our results establish a direct correspondence between local scale symmetry and length-preserving affine geometry, providing a new geometric framework for scale-invariant metric-affine gravity.
\end{abstract}

\maketitle

\section{Introduction}
Metric-affine theories of gravity \cite{Hehl:1994ue,Hehl:1976my,Ivanenko:1983fts,Gronwald:1995em,Obukhov:2002tm,Sotiriou:2006qn,Capozziello:2007tj,Vitagliano:2010sr,Iosifidis:2019dua,Shimada:2018lnm,Percacci:2020bzf} have attracted considerable attention in recent years, since they provide a natural extension of general relativity by promoting the metric and affine connection to independent geometrical variables. Depending on the geometric constraints imposed on the connection, gravity can be formulated in terms of curvature, torsion, or nonmetricity, leading respectively to Riemannian, teleparallel and symmetric teleparallel descriptions (for incomplete references see e.g. \cite{BeltranJimenez:2019esp,Capozziello:2022zzh,Nojiri:2017ncd,Cai:2015emx,Heisenberg:2023lru,Haghani:2012bt,Harko:2011kv,Xu:2019sbp,Sotiriou:2008rp,Bahamonde:2021gfp}). Among these, symmetric teleparallel gravity (STG) \cite{Nester:1998mp,BeltranJimenez:2018vdo,Adak:2005cd,Adak:2008gd,Mol:2014ooa,BeltranJimenez:2017tkd,BeltranJimenez:2019tme}, characterized by vanishing curvature and torsion, attributes the gravitational interaction entirely to nonmetricity and has become an active framework for modified gravity.

A fundamental issue associated with nonmetricity is the second clock effect (SCE). In the original Weyl geometry \cite{weyl1918gravitation, Wheeler:2018rjb,Scholz:2017pfo}, the nonmetricity takes the vectorial form $Q_{\lambda\mu\nu}\equiv\nabla_\lambda g_{\mu\nu}=\omega_\lambda g_{\mu\nu}$, implying that the length of a tangent vector changes under parallel transport. Einstein pointed out that such a path-dependent change of length would contradict the observed stability of atomic spectra \cite{pauli2013theory}, posing a serious challenge to Weyl's original unification program. Recently, some approaches have been proposed to solve the SCE problem, see e.g. \cite{Avalos:2016unj,BeltranJimenez:2020sih,Hobson:2020doi,Pala:2022rry,Quiros:2021eju,Quiros:2022mut}. 
Among these possibilities, Schr\"{o}dinger geometry ~\cite{schrodinger1985space} proposed by Schr\"{o}dinger decades ago offers a particularly elegant and purely geometrical resolution of the SCE.

In recent years, Schr\"{o}dinger geometry has re-emerged as a promising alternative to Weyl geometry for constructing non-Riemannian theories of gravity. By replacing Weyl's metric compatibility condition with the constraint $Q_{(\lambda\mu\nu)}=0$, it preserves the length of tangent vectors under autoparallel transport while retaining a nontrivial nonmetricity. This property naturally avoids the SCE without abandoning the metric-affine framework. Building upon this geometric structure, Schr\"{o}dinger geometry has been extended in several directions, including lower-dimensional gravity \cite{Klemm:2020gfm}, generalized gravity theories \cite{Ming:2023zqv}, length-preserving biconnection gravity \cite{Csillag:2024eor,Csillag:2024bvc} and cosmological models \cite{Chaudhary:2025yvz,Ming:2025mrs}. These developments indicate that
Schr\"{o}dinger geometry is no longer merely
an alternative mathematical construction,
but has evolved into a viable geometrical framework for metric-affine gravity.

Despite these advances, an important question has remained largely unexplored. Local scale symmetry plays a central role in Weyl geometry, where the affine connection is preserved under Weyl rescalings of the metric. However, in the metric-affine formalism the metric and affine connection are independent geometrical variables, and therefore there is no \emph{a priori} reason for the Schr\"{o}dinger affine structure to remain compatible with such transformations \cite{Iosifidis:2018zwo,Iosifidis:2019fsh,Vazirian:2013baa,Sauro:2022hoh,2936998}. 

Understanding how local scale transformations act on the underlying metric-affine geometry is therefore a fundamental issue rather than a model-dependent one. Unlike Weyl geometry, such a scale symmetry, if it exists,
is expected to require simultaneous
transformations of both the metric
and the affine connection. 

This naturally raises the following questions: {\it  does Schr\"{o}dinger geometry admit a local scale symmetry analogous to that of Weyl geometry?} If so, can such a symmetry be realized within symmetric teleparallel gravity? Addressing these questions constitutes the primary objective of the present work.

In this work, we show that a scale invariance exists in the Schr\"{o}dinger geometry. We first determine the corresponding scale transformation compatible with autoparallelism, torsionlessness and Schr\"{o}dinger affine structure in Section \ref{setup}. We then construct a quadratic STG action invariant under the corresponding scale transformation in Section \ref{SIofSTG}. Finally, using the Palatini formalism, in Section \ref{EoMs} we show that the conditions for scale invariance coincide with those required for the field equations to admit a Schr\"{o}dinger connection. This establishes a direct correspondence between local scale symmetry and length-preserving affine geometry in symmetric teleparallel gravity. We discuss and conclude in Section \ref{DisCon}.

\section{Schr\"{o}dinger geometry and scale transformations}\label{setup}
Before constructing a scale-invariant theory, we first identify the geometric properties that characterize a Schr\"{o}dinger connection. These properties will later determine the form of the admissible scale transformation. In the first subsection we revisit the definition for an affine connection to be Schr\"{o}dinger-type, i.e., a connection preserving the length of tangent vectors under the autoparallel transport. We then introduce the scale transformations on the metric and connection under which a theory with nonmetricity preserves autoparallelism, torsionlessness and Schr\"{o}dinger affine structure.

\subsection{Schr\"{o}dinger geometry}
There are three independent geometric variables that can be constructed from the two foundational variables $g_{\mu\nu}$ and $\Gamma^\lambda_{\mu\nu}$, namely curvature,  nonmetricity and torsion,
\begin{align}
    R^{\mu}_{~\nu\alpha\beta}&\equiv\partial_{\alpha}\Gamma^\mu_{~\nu\beta}-\partial_{\beta}\Gamma^\mu_{~\nu\alpha}+\Gamma^\mu_{~\rho\alpha}\Gamma^\rho_{~\nu\beta}-\Gamma^\mu_{~\rho\beta}\Gamma^\rho_{~\nu\alpha},\\
    Q_{\lambda\mu\nu}&\equiv\nabla_\lambda g_{\mu\nu}=\partial_\lambda g_{\mu\nu}-\Gamma^\rho_{~\mu\lambda}g_{\rho\nu}-\Gamma^\rho_{~\nu\lambda}g_{\mu\rho},\\
    T^\lambda_{~[\mu\nu]}&\equiv\Gamma^\lambda_{~[\mu\nu]}=\frac{1}{2}(\Gamma^\lambda_{~\mu\nu}-\Gamma^\lambda_{~\nu\mu}).
\end{align}

Measured by the affine connection $\nabla_\mu$, a curve $C(\tau)$ is autoparallel if its tangent vector $T^\mu\equiv\frac{\partial x^\mu}{\partial\tau}$ satisfies
\begin{equation}
    T^\lambda\nabla_\lambda T^\mu\propto T^\mu,
\end{equation}
and $\tau$ is called an affine parameter if the right hand side is zero. The change of the length of $T^\mu$ under autoparallel transport with affine parameterization (i.e., $T^\lambda\nabla_\lambda T^\mu=0$) is 
\begin{equation}\label{Length Parallel Transport}
\begin{aligned}
  \delta l&=T^\lambda\nabla_\lambda(g_{\mu\nu}T^\mu T^\nu)\\
  &=T^\lambda T^\mu T^\nu \nabla_\lambda g_{\mu\nu}+g_{\mu\nu}T^\lambda T^\nu \nabla_\lambda T^\mu+g_{\mu\nu}T^\lambda T^\mu \nabla_\lambda T^\nu\\
  &=Q_{\lambda\mu\nu}T^\lambda T^\mu T^\nu \\
  &=Q_{(\lambda\mu\nu)}T^\lambda T^\mu T^\nu.
\end{aligned}
\end{equation}
Therefore, only the totally symmetric part of the nonmetricity contributes to the variation of the vector length, and a Schr\"{o}dinger connection defined by $\delta l=0$ requires 
\begin{equation}\label{QtotalSYM}
    Q_{(\lambda\mu\nu)}=0.
\end{equation}
If a Schr\"{o}dinger-type nonmetricity is encoded in a metric-affine theory of gravity as above, the SCE problem does not arise for autoparallel curves.

By contracting with the metric, one obtains the two independent vectorial degrees of freedom of nonmetricity tensor,
\begin{equation}
    \begin{aligned}
    Q_\lambda&\equiv g^{\mu\nu}Q_{\lambda\mu\nu},\\
    \widetilde{Q}_\nu&\equiv g^{\lambda\mu} Q_{\lambda\mu\nu}.
    \end{aligned}
\end{equation}
To make the independent degrees of freedom explicit, we decompose the nonmetricity tensor into its irreducible components,
\begin{equation}\label{Qdecom}
    Q_{\lambda\mu\nu}=\frac{5Q_\lambda-2\widetilde{Q}_\lambda}{18}g_{\mu\nu}+\frac{4\widetilde{Q}_{(\mu}g_{\nu)\lambda}-Q_{(\mu}g_{\nu)\lambda}}{9}+\Omega_{\lambda\mu\nu},
\end{equation}
where $\Omega_{\lambda\mu\nu}$ is the traceless part of $Q_{\lambda\mu\nu}$. For simplicity, we assume that $\Omega_{\lambda\mu\nu}=0$ to deduce the degrees of freedom and focus on the vectorial nonmetricity analogous to Weyl gravity. An equivalent expression for (\ref{QtotalSYM}) from (\ref{Qdecom}) is then obtained as
\begin{equation}\label{SC}
    \widetilde{Q}_\lambda=-\frac{1}{2}Q_\lambda.
\end{equation}
Therefore, the two nonmetricity vectors become dependent on each other and the vectorial Schr\"{o}dinger-type nonmetricity is
\begin{equation}\label{Qqqq}
    Q_{\lambda\mu\nu}=\frac{1}{3}Q_\lambda g_{\mu\nu}-\frac{1}{6}Q_\mu g_{\nu\lambda}-\frac{1}{6}Q_\nu g_{\mu\lambda},
\end{equation}
which clearly satisfies the totally symmetrization condition (\ref{QtotalSYM}).
These relations uniquely characterize the vectorial Schr\"{o}dinger geometry. As a comparison, in Weyl geometry one has $\widetilde{Q}_\lambda=\frac{1}{4}Q_\lambda$, and the nonmetricity tensor is determined by a single independent Weyl vector as 
\begin{equation}
    Q_{\lambda\mu\nu}=\omega_\lambda g_{\mu\nu}=\frac{1}{4}Q_\lambda g_{\mu\nu}=\widetilde{Q}_\lambda g_{\mu\nu}.
\end{equation}  

\subsection{Scale transformations}
Since metric and affine connection are independent variables in metric-affine gravity, a local scale transformation should in general act on both geometrical objects. We therefore consider a Weyl transformation 
\begin{equation}\label{ConfTrans}
    g'_{\mu\nu}=e^\phi g_{\mu\nu}
\end{equation}
on metric and a projective transformation
\begin{equation}
    \label{GammaTrans}\Gamma'^\lambda_{~~\mu\nu}=\Gamma^\lambda_{~\mu\nu}+a\delta^\lambda_\mu \xi_\nu+b\delta^\lambda_\nu \xi_\mu+c g_{\mu\nu}\xi^\lambda
\end{equation}
on connection, where $\phi=\phi(x)$ is an arbitrary scalar function, $\xi_\mu$ a vector function and $a$, $b$, $c$ are undetermined constants.
The most general linear transformation on $\Gamma$ contains three independent vectors \cite{Sauro:2022hoh}. However, as we shall see, {\it preserving the Schr\"{o}dinger affine structure requires only one independent vector.}

We now determine the three constants in (\ref{GammaTrans})  to find the scale transformation that preserve the following properties,
\begin{itemize}
    \item 
    autoparallelism: $T^\lambda\nabla'_\lambda T^\mu\propto T^\mu$ if $T^\lambda\nabla_\lambda T^\mu=0$,
    \item 
    torsionlessness: $T'^\lambda_{~~\mu\nu}=0$ if $T^\lambda_{~~\mu\nu}=0$,
    \item 
    Schr\"{o}dinger structure: $Q'_{(\lambda\mu\nu)}=0$ if $Q_{(\lambda\mu\nu)}=0$.
\end{itemize}
The first one can be evaluate as
\begin{equation}
    \begin{aligned}
&T^\lambda\nabla'_\lambda T^\mu\\
=&T^\lambda\left(\partial_\lambda T^\mu+\Gamma'^\mu_{~~\rho\lambda}T^\rho\right)\\ 
=&T^\lambda\left[\partial_\lambda T^\mu+\left(\Gamma^\mu_{~~\rho\lambda}+a\delta^\mu_\rho\xi_\lambda+b\delta^\mu_\lambda\xi_\rho+cg_{\rho\lambda}\xi^\mu\right)T^\rho\right]\\
=&(a+b)T^\lambda \xi_\lambda T^\mu+cT^\lambda T_\lambda \xi^\mu,
    \end{aligned}
\end{equation}
and thus preserving autoparallelism requires $c=0$. For the condition of torsionlessness, one finds $a=b$. For the connection to preserve Schr\"{o}dinger affine structure under (\ref{ConfTrans}) and (\ref{GammaTrans}) with the previous two conditions satisfied, we have
\begin{equation}\label{eq15}
    \begin{aligned}
        0=&Q'_{(\lambda\mu\nu)}\\
        =&\partial_{(\lambda} g'_{\mu\nu)}-\Gamma'^\rho_{~~(\mu\lambda}g'_{\nu)\rho}-\Gamma'^\rho_{~~(\nu\lambda}g'_{\mu)\rho}\\
        =&\partial_{(\lambda}\left( e^\phi g_{\mu\nu)} \right)-\left( \Gamma^\rho_{~(\mu\lambda}+a\delta^\rho_{(\mu}\xi_\lambda+a\delta^\rho_{(\lambda}\xi_\mu \right)e^\phi g_{\nu)\rho}\\
        &-\left( \Gamma^\rho_{~(\nu\lambda}+a\delta^\rho_{(\nu}\xi_\lambda+a\delta^\rho_{(\lambda}\xi_\nu \right)e^\phi g_{\mu)\rho}\\
        =&e^\phi\left[ Q_{(\lambda\mu\nu)}+g_{(\mu\nu}\left( \partial_{\lambda)}\phi-4a\xi_{\lambda)} \right) \right]\\
        =&e^\phi g_{(\mu\nu}\left( \partial_{\lambda)}\phi-4a\xi_{\lambda)} \right),
    \end{aligned}
\end{equation}
therefore we get 
\begin{equation}
    a\xi_\lambda=\frac{1}{4}\partial_\lambda \phi.
\end{equation}

In a short summary, we will consider the scale transformation,
\begin{equation}\label{ST}
\begin{aligned}
    g'_{\mu\nu}&=e^\phi g_{\mu\nu},\\
    \Gamma'^\lambda_{~~\mu\nu}&=\ \Gamma^\lambda_{~~\mu\nu}+\frac{1}{4}\delta^\lambda_\mu\partial_\nu\phi+\frac{1}{4}\delta^\lambda_\nu\partial_\mu\phi,
\end{aligned}
\end{equation}
under which autoparallelism, torsionlessness and Schr\"{o}dinger affine structure are preserved. The projective transformation on the connection coefficients is the analogue of $\Gamma'^\lambda_{~~\mu\nu}=\Gamma^\lambda_{~~\mu\nu}$ in Weyl geometry, and it is equivalent to $Q'_\lambda=Q_\lambda+\frac{3}{2}\partial_\lambda\phi$.

In the following  we shall refer to (\ref{ST}) as the Schr\"{o}dinger scale transformation. One can check that the nonmetricity tensor transforms as
\begin{equation}
    Q'_{\lambda\mu\nu}=\frac{1}{3}Q'_\lambda g'_{\mu\nu}-\frac{1}{6}Q'_\mu g'_{\nu\lambda}-\frac{1}{6}Q'_\nu g'_{\mu\lambda}
\end{equation}
from its definition and the Schr\"{o}dinger scale transformation.

Further, if the tangent vector is required to be affine parameterized evaluated by $\nabla'$ so that it corresponds to the physical $4$-velocity along a worldline also after the Schr\"{o}dinger scale transformation, i.e., $T'^\lambda\nabla'_\lambda T'^\mu=0$, its transformation must be $T'^\mu=e^{-\frac{1}{2}\phi}T^\mu$. To check this, 
we note
\begin{equation}
	\begin{aligned}
        T'^\lambda\nabla'_\lambda T'^\mu =& e^{-\frac{1}{2}\phi}T^\lambda\nabla'_\lambda \left( e^{-\frac{1}{2}\phi}T^\mu \right) \\
        =&e^{-\frac{1}{2}\phi}T^\mu T^\lambda\nabla'_\lambda e^{-\frac{1}{2}\phi} + e^{-\phi}T^\lambda\nabla'_\lambda T^\mu \\
        =&e^{-\frac{1}{2}\phi}T^\mu T^\lambda\partial_\lambda e^{-\frac{1}{2}\phi} + e^{-\phi}T^\lambda\partial_\lambda T^\mu\\
        &+e^{-\phi}T^\lambda T^\rho \left( \Gamma^\mu_{~\rho\lambda}+\frac{1}{4}\delta^\mu_{\rho}\partial_\lambda\phi+\frac{1}{4}\delta^\mu_{\lambda}\partial_\rho\phi \right) \\
        =&0. 
	\end{aligned}
\end{equation}

Consequently, the weight of $4$-velocity vector being $-\frac{1}{2}$, leads to the result that the norm of the tangent vector is scale invariant,
\begin{equation}
	g'_{\mu\nu}T'^\mu T'^\nu=g_{\mu\nu}T^\mu T^\nu.
\end{equation}

As a crosscheck, the length change of the tangent vector under autoparallel transport, $\delta l$, transforms covariantly,
\begin{equation}
\begin{aligned}
\delta l'=&T'^\lambda\nabla'_\lambda \left( g'_{\mu\nu}T'^\mu T'^\nu \right)\\
=&T'^\lambda T'^\mu T'^\nu \nabla'_\lambda g'_{\mu\nu}+g'_{\mu\nu}T'^\lambda T'^\nu \nabla'_\lambda T'^\mu+g'_{\mu\nu}T'^\lambda T'^\mu \nabla'_\lambda T'^\nu\\
=&Q'_{(\lambda\mu\nu)}T'^\lambda T'^\mu T'^\nu\\
=&e^\phi Q_{(\lambda\mu\nu)}e^{-\frac{3}{2}\phi}T^\lambda T^\mu T^\nu\\
=&e^{-\frac{1}{2}\phi}\delta l.
\end{aligned}
\end{equation}
Thus, $\delta l=0$ implies $\delta l'=0$, the Schr\"{o}dinger affine structure is preserved  and SCE will not appear due to the Schr\"{o}dinger scale transformation, as already shown in (\ref{eq15}). 

For convenience, the main differences between Weyl and Schr\"{o}dinger geometries, together with their corresponding scale transformations, are summarized in Table \ref{table}.

\renewcommand{\arraystretch}{1.5}
\begin{table*}[]
    \centering
    \begin{tabular}{|c|c|c|}
    \hline \hline
         Geometry & Weyl & Schr\"{o}dinger \\
         \hline\hline
         $Q_{\lambda\mu\nu}$ & $\frac{1}{4}Q_\lambda g_{\mu\nu}$ & $\frac{1}{3}Q_\lambda g_{\mu\nu}-\frac{1}{6}Q_\mu g_{\nu\lambda}-\frac{1}{6}Q_\nu g_{\mu\lambda}$ \\
         \hline
         $Q_{(\lambda\mu\nu)}$ & $\neq 0$ & $0$ \\
         \hline
         $\delta l$ & $\neq 0$ & 0 \\
         \hline
         $\widetilde{Q}_\lambda$ & $\frac{1}{4}Q_\lambda$ & $-\frac{1}{2}Q_\lambda$ \\
         \hline
         $\Gamma^\lambda_{~\mu\nu}$ & $\{^\lambda_{\mu\nu}\}+\frac{1}{8}Q^\lambda g_{\mu\nu}-\frac{1}{8}Q_\mu\delta^\lambda_\nu-\frac{1}{8}Q_\nu\delta^\lambda_\mu$ & $\{^\lambda_{\mu\nu}\}+\frac{1}{3}Q^\lambda g_{\mu\nu}-\frac{1}{6}Q_\mu\delta^\lambda_\nu-\frac{1}{6}Q_\nu\delta^\lambda_\mu$ \\
         \hline
         \hline
         $g'_{\mu\nu}$ & $e^\phi g_{\mu\nu}$ & $e^\phi g_{\mu\nu}$ \\
         \hline
         $\Gamma'^\lambda_{~~\mu\nu}$ & $\Gamma^\lambda_{~~\mu\nu}$ & $\Gamma^\lambda_{~~\mu\nu}+\frac{1}{4}\delta^\lambda_\mu\partial_\nu\phi+\frac{1}{4}\delta^\lambda_\nu\partial_\mu\phi$ \\
         \hline
         $Q'_{\lambda\mu\nu}$ & $e^\phi\left(Q_{\lambda\mu\nu}+g_{\mu\nu}\partial_\lambda\phi\right)$ &  $e^\phi \left( Q_{\lambda\mu\nu} + \frac{1}{2}g_{\mu\nu}\partial_\lambda\phi - \frac{1}{4}g_{\lambda\nu}\partial_{\mu}\phi - \frac{1}{4}g_{\mu\lambda}\partial_\nu\phi \right)$ \\
          & =$\frac{1}{4}Q'_\lambda g'_{\mu\nu}$ & =$\frac{1}{3}Q'_\lambda g'_{\mu\nu}-\frac{1}{6}Q'_\mu g'_{\nu\lambda}-\frac{1}{6}Q'_\nu g'_{\mu\lambda}$ \\
         \hline
         $Q'_\lambda$ & $Q_\lambda+4\partial_\lambda\phi$ & $Q_\lambda+\frac{3}{2}\partial_\lambda\phi$\\
         \hline \hline
    \end{tabular}
    \caption{Comparison between Weyl and Schr\"{o}dinger geometries and their corresponding local scale transformations. The Schr\"{o}dinger scale transformation preserves the length-preserving affine structure while inducing transformation laws analogous to those in Weyl geometry. Here we present the results in terms of the vector $Q_\lambda$, which is connected with the Weyl potential $\omega_\lambda$ (defined by $Q_\lambda=\omega_\lambda g_{\mu\nu}$) via  $Q_\lambda=4\omega_\lambda$.}
    \label{table}
\end{table*}

\section{Construction of a scale-invariant quadratic STG theory}\label{SIofSTG}
Having determined the Schr\"{o}dinger scale transformation, we now ask whether a symmetric teleparallel action can respect this symmetry. 

The gravitational action in STG is usually constructed from the nonmetricity scalar $\mathbb{Q}$ \cite{Heisenberg:2023lru}
\begin{equation}
    \begin{aligned}\mathbb{Q}=&c_1Q_{\lambda\mu\nu}Q^{\lambda\mu\nu}+c_2Q_{\lambda\mu\nu}Q^{\mu\lambda\nu}+c_3Q_\lambda Q^\lambda\\
    &+c_4\tilde{Q}_\lambda\tilde{Q}^\lambda+c_5Q_\lambda\tilde{Q}^\lambda,
    \end{aligned}
\end{equation}
where $c_1,\dots,c_5$ are undetermined coefficients. Since the nonmetricity tensor $Q_{\lambda\mu\nu}$ acquires a factor $e^\phi$ under Weyl transformation, $\mathbb{Q}$ transforms with weight $-1$. While $\sqrt{-g}$ carries weight $+2$, a linear action in $\mathbb{Q}$ cannot be scale invariant without introducing additional fields. 

Therefore, to study a theory of STG with all gravitational information encoded in the nonmetricity tensor, we consider the quadratic action of symmetric teleparallel gravity, 
\begin{equation}\label{action}
		S=\int d^4x \sqrt{-g}\mathbb{Q}^2.
\end{equation}

Under the Schr\"{o}dinger scale transformation, $Q_{\lambda\mu\nu}$ and its two independent vectorical degrees of freedom transform as
\begin{equation}
	\begin{aligned}
        Q'_{\lambda\mu\nu} &= e^\phi \left( Q_{\lambda\mu\nu} + \frac{1}{2}g_{\mu\nu}\partial_\lambda\phi - \frac{1}{4}g_{\lambda\nu}\partial_{\mu}\phi - \frac{1}{4}g_{\mu\lambda}\partial_\nu\phi \right), \\
        Q'_\lambda &= g'^{\mu\nu}Q'_{\lambda\mu\nu} = Q_\lambda + \frac{3}{2}\partial_\lambda\phi, \\
        \tilde{Q}'_\nu &= g'^{\lambda\mu}Q'_{\lambda\mu\nu} = \tilde{Q}_\nu - \frac{3}{4}\partial_\nu\phi.
	\end{aligned}
\end{equation}
We note that the transformation rule on $Q_\lambda$ or $\widetilde{Q}_\lambda$ is the analogue of the gauge transformation $\omega'_\lambda=\omega_\lambda+\partial_\lambda\phi$ in Weyl geometry. The nonmetricity scalar transforms as
\begin{equation}
	\begin{aligned}
        \mathbb{Q}'=&c_1Q'_{\lambda\mu\nu}Q'^{\lambda\mu\nu}+c_2Q'_{\lambda\mu\nu}Q'^{\mu\lambda\nu}+c_3Q'_\lambda Q'^\lambda\\
        &+c_4\tilde{Q}'_\lambda\tilde{Q}'^\lambda+c_5Q'_\lambda\tilde{Q}'^\lambda\\
        =&e^{-\phi} \left[ \mathbb{Q} + \left( c_1-\frac{1}{2}c_2+3c_3-\frac{3}{4}c_5 \right) Q^\rho\partial_\rho\phi \right. \\
        &\left. + \left( -c_1+\frac{1}{2}c_2-\frac{3}{2}c_4+\frac{3}{2}c_5 \right) \tilde{Q}^\rho\partial_\rho\phi \right. \\
        &\left. + \left( \frac{9}{8}c_1-\frac{9}{16}c_2+\frac{9}{4}c_3+\frac{9}{16}c_4-\frac{9}{8}c_5 \right) \partial^\rho\phi\partial_\rho\phi \right].
	\end{aligned}
\end{equation}
	It is clear that if the coefficients $c_i$ satisfy the conditions
\begin{equation}\label{constant_1}
	\begin{aligned}
		&c_1-\frac{1}{2}c_2+3c_3-\frac{3}{4}c_5=0,\\
		&-c_1+\frac{1}{2}c_2-\frac{3}{2}c_4+\frac{3}{2}c_5=0,\\
		&2c_1-c_2+4c_3+c_4-2c_5=0,
	\end{aligned}
\end{equation}
    the nonmetricity scalar transforms covariantly,
\begin{equation}
    \mathbb{Q}'=e^{-\phi}\mathbb{Q}.
\end{equation}
These three conditions are both necessary and sufficient for the quadratic action (\ref{action}) to be scale invariant,
    \begin{equation}
	S'=\int d^4x \sqrt{-g'}\mathbb{Q}'^2=S.
    \end{equation}

\section{Dynamical emergence of the Schr\"{o}dinger geometry}
\label{EoMs}
We now investigate whether the Schr\"{o}dinger geometry also emerges dynamically from the field equations. We shall study the equations of motion for (\ref{action}) based on the Palatini formalism, in which metric and connection are regarded as two independent geometrical variables and the action should be varied with respect to both. 

By performing the variation of the action with respect to the connection $\Gamma^\lambda_{~\mu\nu}$ while keeping the metric fixed, we find
\begin{equation}
	\begin{aligned}
		\delta_\Gamma S=&\int d^4x\sqrt{-g}\delta_{\Gamma}\mathbb{Q}^2\\
		=&\int d^4x2\sqrt{-g}\mathbb{Q}\delta_{\Gamma}\mathbb{Q}\\
		=&\int d^4x\sqrt{-g}\left( -8P^{\nu\mu}_{~~~\lambda}\mathbb{Q} \right)\delta\Gamma^{\lambda}_{~\mu\nu},
	\end{aligned}
\end{equation}
where $P^{\nu\mu}_{~~~\lambda}$ is the so-called nonmetricity conjugate defined as
\begin{equation}\label{Ptensor}
    \begin{aligned}
        P^{\alpha\beta\gamma}=&c_1Q^{\alpha\beta\gamma}+c_2Q^{(\beta|\alpha|\gamma)}+c_3Q^\alpha g^{\beta\gamma} \\
        &+c_4g^{\alpha(\beta}\tilde{Q}^{\gamma)}+\frac{c_5}{2} \left( \tilde{Q}^\alpha g^{\beta\gamma}+g^{\alpha(\beta}Q^{\gamma)} \right),
    \end{aligned}
\end{equation}
and one obtains the field equation for connection
\begin{equation}
    P^{\nu\mu}_{~~~\lambda}=0.
\end{equation}
Thus, the connection equation is purely algebraic in terms of the two vectorial traces of the nonmetricity. By substituting the trace decomposition of nonmetricity (\ref{Qdecom}) into (\ref{Ptensor}), one obtains
\begin{equation}\label{P}
	\begin{aligned}
        &\left( -\frac{1}{18}c_1+\frac{1}{9}c_2+\frac{1}{4}c_5 \right) Q_\lambda g^{\mu\nu} + \left( \frac{2}{9}c_1+\frac{1}{18}c_2+\frac{1}{2}c_4 \right) \\
        &\tilde{Q}_\lambda g^{\mu\nu} + \left( -\frac{1}{18}c_1+\frac{1}{9}c_2+\frac{1}{4}c_5 \right) Q^{\mu}\delta^{\nu}_{\lambda}+ \left( \frac{2}{9}c_1+\frac{1}{18}c_2\right. \\
        &\left.+\frac{1}{2}c_4 \right) \tilde{Q}^{\mu}\delta^{\nu}_{\lambda} + \left( \frac{5}{18}c_1-\frac{1}{18}c_2+c_5\right)Q^{\nu}\delta^{\mu}_{\lambda}+\\
        &\left( -\frac{1}{9}c_1+\frac{2}{9}c_2+\frac{1}{2}c_5 \right) \tilde{Q}^{\nu}\delta^{\mu}_{\lambda}+\Omega^{\nu\mu}_{~~\lambda}+\Omega^{\mu\nu}_{~~\lambda}+\Omega^{\lambda}_{~\mu\nu}=0.
	\end{aligned}
\end{equation}

Taking the independent traces by contracting (\ref{P}) with $g_{\mu\nu}, \delta^{\lambda}_{\mu}$ and $\delta^{\lambda}_{\nu}$ respectively, the tensorial components are eliminated and it yields algebraic relations for the vectorial degrees of freedom,
\begin{equation}\label{eq33}
	\begin{aligned}
        &\left( \frac{1}{2}c_2 + c_3 + \frac{5}{4}c_5 \right) Q_{\lambda} + \left( c_1 + \frac{1}{2}c_2 + \frac{5}{2}c_4 + \frac{1}{2}c_5 \right) \tilde{Q}_\lambda = 0, \\
        &\left( c_1 + 4c_3 + \frac{1}{2}c_5 \right) Q^\nu + \bigg( c_2 + c_4 + 2c_5 \bigg) \tilde{Q}^\nu = 0.
	\end{aligned}
\end{equation}
If the connection is required to be Schr\"{o}dinger-type and vectorial, i.e., (\ref{SC}) holds, one arrives at another set of equations for the coefficients from (\ref{eq33})
\begin{equation}\label{constant_2}
	\begin{aligned}
		&c_1-\frac{1}{2}c_2-2c_3+\frac{5}{2}c_4-2c_5=0,\\
		&2c_1-c_2+8c_3-c_4-c_5=0.
	\end{aligned}
\end{equation}

Remarkably, the conditions required for scale invariance coincide exactly with those ensuring that the connection is dynamically reduced to Schr\"{o}dinger form. (\ref{constant_1}) and (\ref{constant_2}) are indeed linearly dependent on each other and admit same solutions characterized by three free parameters out of five $c_i$,
\begin{equation}\label{resultsOFci}
	\begin{aligned}
		&c_4=\frac{2}{3}c_1-\frac{1}{3}c_2+4c_3,\\
		&c_5=\frac{4}{3}c_1-\frac{2}{3}c_2+4c_3.
	\end{aligned}
\end{equation}
This is not accidental. (\ref{constant_1}) is obtained by requiring the action (\ref{action}) to be invariant under the Schr\"{o}dinger scale transformation (\ref{ST}), while (\ref{constant_2}) leads to that a Schr\"{o}dinger connection with (\ref{SC}) is automatically fixed by (\ref{action}). Rather, it reflects the fact that the Schr\"{o}dinger geometry is scale invariant in the quadratic STG: if the geometry is Schr\"{o}dinger before the scale transformation, it will still be after, since the Schr\"{o}dinger scale transformation (\ref{ST}) preserves Schr\"{o}dinger affine structure.

For completeness, we perform the variation of action with respect to the inverse metric while keeping the connection fixed,

\begin{equation}
    \begin{aligned}
      \delta_gS =&\int d^4x \delta_g \left( \sqrt{-g}\mathbb{Q}^2 \right) \\
        =&\int d^4x \left( 2\mathbb{Q}\sqrt{-g}\delta_g\mathbb{Q} - \frac{1}{2}\mathbb{Q}^2\sqrt{-g}g_{\alpha\beta}\delta g^{\alpha\beta} \right) \\
        =&\int d^4x \sqrt{-g}\left[\frac{4}{\sqrt{-g}}\nabla_\lambda\left(\sqrt{-g}\mathbb{Q}P^\lambda_{~\alpha\beta}\right)-2\mathbb{Q}q_{\alpha\beta} \right.\\
        &\left.-\frac{1}{2}\mathbb{Q}^2g_{\alpha\beta}\right]\delta g^{\alpha\beta},
    \end{aligned}
\end{equation}
where
\begin{equation}
\begin{aligned}
     q_{\mu\nu}\equiv&  c_1 \left(2Q_{\mu\nu\alpha} Q^{\mu\nu}{}_\beta- Q_{\alpha\mu\nu} Q_\beta{}^{\mu\nu} \right)+ c_2 Q_{\mu\nu\alpha} Q^{\nu\mu}{}_\beta \\
        & + c_3 \left(2Q^\lambda Q_{\lambda\alpha\beta}- Q_\alpha Q_\beta \right)+c_4 \tilde{Q}_\alpha \tilde{Q}_\beta+c_5 \tilde{Q}^\lambda Q_{\lambda\alpha\beta}.
\end{aligned}
\end{equation}

Finally, one can read off the dynamic field equation for the metric as
\begin{equation}
        \frac{4}{\sqrt{-g}}\nabla_\lambda\left(\sqrt{-g}\mathbb{Q}P^\lambda_{~\alpha\beta}\right)-2\mathbb{Q}q_{\alpha\beta}-\frac{1}{2}\mathbb{Q}^2g_{\alpha\beta}=0
\end{equation}

\section{Conclusion and discussion}
\label{DisCon}
In this work, we have investigated the realization of local scale symmetry in Schr\"{o}dinger geometry within the framework of symmetric teleparallel gravity. We first determined the unique scale transformation (\ref{ST}) that preserves the defining properties of Schr\"{o}dinger geometry, namely autoparallelism, torsionlessness and length preservation under autoparallel transport. Based on this  transformation, we constructed a quadratic symmetric teleparallel action that is scale invariant. 

Furthermore, by employing the Palatini formalism, we showed that the conditions (\ref{constant_1}), required for scale invariance, coincide exactly with (\ref{constant_2}) under which the affine connection is dynamically reduced to the Schr\"{o}dinger form.

The equivalence between the symmetry conditions and the dynamical conditions is the central result of this work. It demonstrates that the Schr\"{o}dinger affine structure is not merely compatible with local scale symmetry, but is naturally selected by the dynamics of the scale-invariant quadratic theory with (\ref{resultsOFci}). In this sense, local scale symmetry and length-preserving affine geometry provide two complementary descriptions of the same underlying geometric structure within symmetric teleparallel gravity. This correspondence points toward a deeper interplay between local scale symmetry and affine geometry, which may provide a useful guiding principle for extending non-Riemannian gravity beyond the Weyl framework.

Several issues deserve further investigation. Our demonstration of the exact correspondence between scale invariance and dynamical Schr\"{o}dinger reduction relies on the vectorial restriction 
$\Omega_{\lambda\mu\nu}=0$. Extending this correspondence to the full nonmetricity tensor with nonvanishing traceless part would require incorporating additional tensor modes into the action and analyzing their scale transformations and field equations. We leave this generalization for future work. The scale transformation derived in this work preserves the defining properties of Schr\"{o}dinger geometry, but does not in general preserve the vanishing Riemann or Ricci curvature. More general linear transformations of the affine connection likewise fail to maintain the flatness condition because additional constraints involving second derivatives of the local scale function $\phi$ inevitably arise. A fully systematic treatment should therefore incorporate the curvature-free and torsion-free conditions into the action through Lagrange multipliers. In such a formulation, the affine connection would become a genuinely dynamical variable, and it would be interesting to investigate whether the Schr\"{o}dinger geometry can emerge dynamically without being imposed as an a priori geometric assumption.

\section*{Acknowledgments}
The work of H.H.Z. was supported by the National Natural Science Foundation of China (NSFC) under Grant No. 12275367, the Fundamental Research Funds for the Central Universities, and the Sun Yat-Sen University Science Foundation.

\bibliography{bib.bib}

\end{document}